# EXTRACTING SUPERCONDUCTING PARAMETERS FROM SURFACE RESISTIVITY BY USING INSIDE TEMPERATURES OF SRF CAVITES*

M. Ge[#], G. Hoffstaetter, H. Padamsee, V. Shemelin
Cornell University, Ithaca, NY, 14853, USA

*Abstract*

The surface resistance $R_s$ of an RF superconductor depends on the surface temperature $T_{in}$, the residual resistance $R_{res}$ and various superconductor parameters, *e.g.* the energy gap, and the electron mean free path. These parameters can be determined by measuring the quality factor $Q_0$ of a SRF cavity in helium-baths of different temperatures. The surface resistance can be computed from $Q_0$ for any cavity geometry, but it is not trivial to determine the temperature $T_{in}$ of the surface when only the temperature of the helium bath is known.

Traditionally, it was approximated that the surface temperature on the inner surface of the cavity was the same as the temperature of the helium bath. This is a good approximation at small RF-fields on the surface, but to determine the field dependence of $R_s$, one cannot be restricted to small field losses.

Here we show the following: (1) How computer simulations can be used to determine the inside temperature $T_{in}$ so that $R_s(T_{in})$ can then be used to extract the superconducting parameters. The computer code combines the well-known programs, the HEAT code and the SRIMP code. (2) How large an error is created when assuming the surface temperature is same as the temperature of the helium bath? It turns out that this error is at least 10% at high RF-fields in typical cases.

## INTRODUCTION

The surface resistance $R_s$ of superconductors under an RF-field is a function of the temperature on the RF surface. In the case of a standing-wave resonator, this is the inner surface of an SRF cavity. To determine $R_s$, the quality factor $Q_0$ has traditionally been measured in a series of the helium-bath temperatures at low RF-fields, usually around 3-5 MV/m. It was typically approximated that the inner temperature was equal to the bath temperature at these low fields. By fitting superconductor parameters in an equation for $R_s(T_{in})$ to the data, quantities like the energy gap, the residual resistance, the electron mean free path *etc.* have been obtained.

Superconductivity theories [1-4] describe the performance of superconductors under low magnetic-field ($H_{RF} \ll H_c$), whether it can be extended to the high RF-fields is unclear [5, 6]. In [7], an effort has been made to empirically establish the relation between the superconducting parameters and the magnetic field by fitting $R_s(T_{in})$ curves to data that was obtained with high RF-fields. This was done up to 100-120mT on the surface, which corresponds to 25-30MV/m accelerating gradient. For such high fields, it is no longer accurate to approximate the inner temperature by the bath temperature.

This paper demonstrates the method of computing the inner temperature from the bath temperature and the RF-field so that the correct relation between the surface resistance and the inner temperature can be made at each RF-field, i.e. calculating $R_s(T_{in})$ from $R_s(T_{bath}, H_p)$. Hence the valid superconducting parameters versus the magnetic fields can be obtained from the $R_s(T_{in})$ curves.

This method converts the surface resistance $R_s$ from the quality factor $Q_0$ by using the geometry factor $G$. $R_s(T_{in})$ relates the magnetic-field distribution on the surface, thus it causes errors of obtaining $R_s$ when $G$ is applied. It has been proved in APPENDIX I that the errors are tiny for elliptical cavities where the distribution of the magnetic field is nearly a constant in the equator region and decays rapidly to the iris region [10]. The proposed method can therefore be used for the TESLA cavity [8], or the Cornell ERL cavity [9]. The method cannot be applied for the high-field Q-slope region where losses are often concentrated in small regions of high surface resistance [11, 12]; it can also not be applied to cavities with large field-emission where losses are not only from dissipation but also due to x-ray radiation.

This paper simulates the baked and the unbaked cases with the energy gap 1.35-1.51 (eV×10⁻³), corresponding to $\frac{\Delta}{kT_c} =$ 1.7-1.9,[1] which covers the most common cases. It turns out that in typical cases of the high-gradient region, the superconductor parameters obtained by fitting $R_s(T_{in})$ differ by at least 10% from those obtained by the traditional method of approximating $T_{in}$ as the bath temperature. In the high-field region it can therefore often be important to apply the here presented, improved method.

## THE FITTING METHODS OF $R_S(T)$ CURVES

The surface resistance of the superconductors under the RF-fields includes two parts, one is the BCS resistance $R_{BCS}$ and the other part is the residual resistance $R_0$ showed in Eq. (1). Eq. (2) is the approximate expression of the surface resistance from the BCS theory [5].

$$R_s = R_{BCS} + R_0 \qquad (1)$$

$$R_{BCS} = A\left(\frac{1}{T_{in}}\right)f^2 \exp\left(-\frac{\Delta}{kT_{in}}\right) \qquad (2)$$

Here, $T_{in}$ is the temperature on inner surface, the factor $A$ is a constant which is determined by material properties

---

*Work supported by NSF award PHY-0969959 and DOE award DOE/SC00008431
#mg574@cornell.edu

[1] $\frac{\Delta}{kT_c}$ is used to express the energy gap in this paper.

*e.g.* the electron mean free path $l_e$ *etc.*; $\Delta$ is the energy gap; *f* is the resonant frequency of the cavity. Eq. (1) and (2) have been widely used for the $R_s(T_{bath})$ curve fittings at low fields to extract the residual resistance.

A better fitting method is based on the SRIMP code. The SRIMP code which incorporates the full BCS theory was written by Jurgen Halbritter [13, 14] for the BCS resistance calculation. Five material parameters are required to describe the superconductor:

- The transition temperature $T_c$;
- The energy gap at $T = 0K$ (entered as $\frac{\Delta}{kT_c}$);
- The London penetration depth at $T = 0K$, $\lambda_0$;
- The coherence length at $T = 0K$, $\xi_0$;
- The electron mean free path at $T = 4.2K$, $l_e$.

The advantage of the SRIMP fitting over the Eq. (2) based method is the SRIMP method is able to describe the properties of superconductors precisely, *e.g.* fitting out the parameter $l_e$ which is related to the 120°C baking. The baked cavity has smaller $l_e$ value than the unbaked cavity [5].

In fact, the fitting methods introduced above don't consider the temperature difference between the inside and the bath temperature. The temperature difference relates to the RF-power, hence it can be written as a function of the peak magnetic fields $H_p$ and the surface resistance $R_s(T_{in})$ for the elliptical cavity. The ratio of the peak magnetic field $H_p$ and the accelerating gradient $E_{acc}$ is a constant, so we use $E_{acc}$ to replace $H_p$, which is given in Eq. (3).

$$T_{in} - T_{bath} = f(R_s(T_{in}), E_{acc}) \quad (3)$$

The bath temperature is measured from experiments. Figure 1, 2 show the procedures of converting the $Q_0$-$E_{acc}$ curves to the $R_s(T_{bath})$ curves as an example. In Figure 1, the temperature listed on right is the helium-bath temperature. The dash line in the figure marks the $Q_0$ values at 20MV/m from the bath temperature 1.7K to 2K. It is possible to convert the $Q_0$ value to $R_s(T_{bath})$ by Eq. (4),

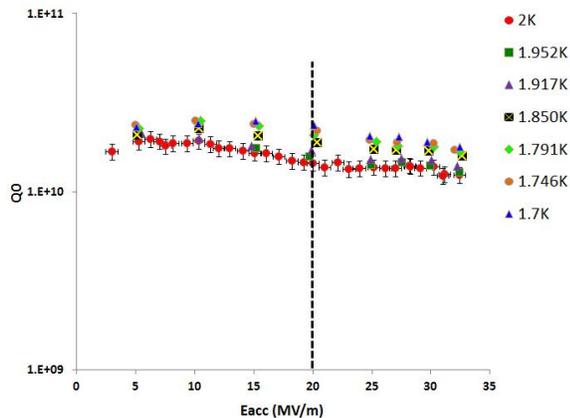

Figure 1: An example of converting $Q_0$-$E_{acc}$ curves to an $R_s$-$1/T_{bath}$ curve at $E_{acc}$=20MV/m.

$$R_s(T_{bath}) = \frac{G}{Q_0(T_{bath})} \quad (4)$$

Here the constant *G* is the geometry factor determined by the cavity shape. The *G* value we used here is 278Ω which is the value of the TESLA cavity. Then it is possible to re-plot the $R_s(T_{bath})$ curves which are shown in Figure 2. By the same way, the $R_s(T_{bath})$ curves at each accelerating gradient are possible to be plotted.

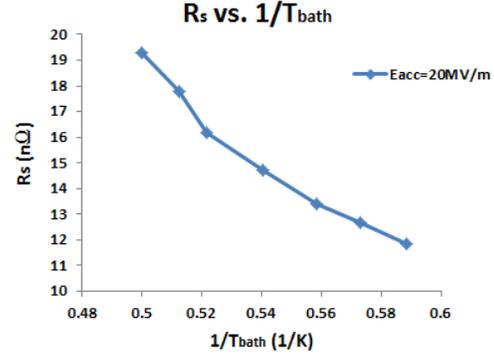

Figure 2: An example of the $R_s$-$1/T_{bath}$ curve converted from Figure 1 at 20MV/m.

## THE HEAT-SRIMP FITTING METHODE

### The temperature rising on interior wall

It is important to calculate the temperature on the inner surface from the bath temperature. The thermal feedback model has been adopted for the calculation [5, 16]. The numerical programs, the HEAT code [15, 16] as well as the HEAT-and-SRIMP program [17], based on this model have been developed at Cornell University. Figure 3 compares the inner temperatures with the bath temperature versus $E_{acc}$, which is calculated by the HEAT and SRIMP program. The program simulated a 1.3GHz cavity (the TESLA cavity) in 2K helium bath with the energy gap $\frac{\Delta}{kT_c}$ varying from 1.7 to 1.9. The cavity tests statistic in Cornell University indicates that $\frac{\Delta}{kT_c}$ of the most cavities is around 1.8-1.9. Based on the BCS theory, the larger energy gap gives the smaller $R_{BCS}$. The 120°C baking of cavities reduces the electron-mean-free-path $l_e$ to achieve smaller $R_{BCS}$ [5]; and it changes the surface of the cavities from the clean limits to the dirty limits. Here we set $l_e$=300Å to represent the baked case shown in Figure 3(a); and $l_e$=6000Å is the unbaked case (Figure 3(b)). The calculation clearly suggests that the temperature increase is more than 0.1K at the accelerating gradient 38MV/m in the baked case with $\frac{\Delta}{kT_c}$ 1.7-1.9 which we define as the typical cases.

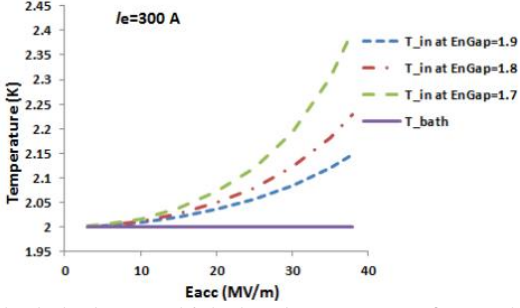

(a) The baked case which the electron mean free path $l_e$ is 300Å.

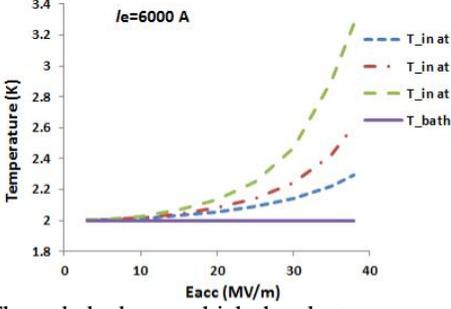

(b) The unbaked case which the electron mean free path $l_e$ is 6000Å.

Figure 3: The temperature increases on inner surface of a 1.3GHz cavity with the energy gap $\frac{\Delta}{kT_c}$ varying from 1.7 to 1.9.

The inner-temperature rise causes the BCS resistance growth when increasing $E_{acc}$, which is depicted in Figure 4 (a) and (b) corresponding to Figure 3 (a) and (b) respectively.

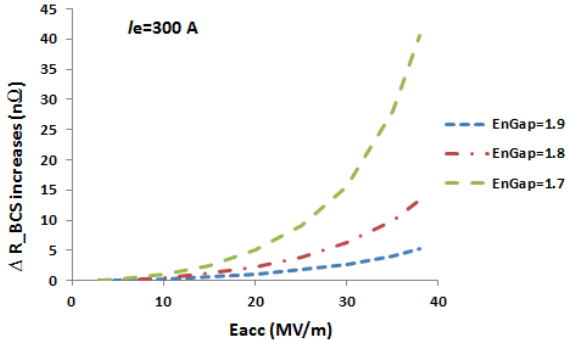

(a) The baked case in which the electron mean free path $l_e$ is 300Å.

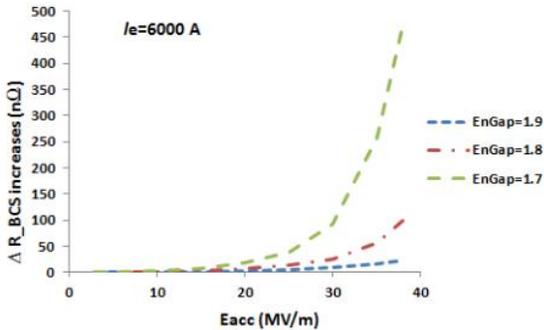

(b) The unbaked case in which the electron mean free path $l_e$ is 6000Å

Figure 4: The BCS resistance increase of a 1.3GHz cavity with $\frac{\Delta}{kT_c}$ varying from 1.7 to 1.9 which is caused by the temperature rising shown in Figure 3 (a) and (b) respectively.

Figure 4(a) shows the baked case; and the increase of the BCS resistance at 38 MV/m is from 5 to 40 nΩ corresponding to $\frac{\Delta}{kT_c}$ from 1.9 to 1.7. Figure 4(b) is the unbaked case; the BCS resistance rising is 23-473 nΩ within the range 1.9-1.7.

The thermal effects cause different Q-slopes in each case exhibited in Figure 5. The six cases covered the most situations of the real measurements; and the fitting error is different in each case.

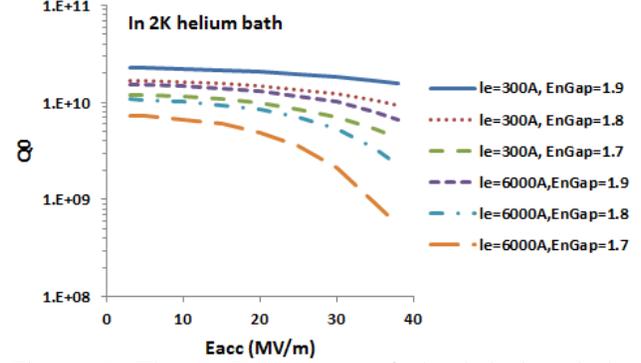

Figure 5: The $Q_0$-$E_{acc}$ curves of the baked and the unbaked cases for $\frac{\Delta}{kT_c}$ 1.7-1.9.

### The HEAT-SRIMP fitting

This section describes how the inner temperature is calculated from the bath temperature; and how to obtain the superconducting parameters by the fitting. We developed the HEAT-SRIMP fitting program which combines the SRIMP code, the HEAT code, and a least square fitting program together.

Here we use the SRIMP code to calculate the surface resistance based on the full BCS theory. The code only considers the linear BCS resistance model. Except the inner-surface temperature $T_{in}$, the rest input-parameters of the SRIMP code are listed in Table 1,

Table 1: The SRIMP input parameters

| |
|---|
| The transition temperature, $T_c$ |
| The cavities resonant frequency, $f_0$ |
| The energy gap at $T = 0K$, $\frac{\Delta}{kT_c}$ |
| The London penetration depth at $T = 0K$, $\lambda_0$ |
| The coherence length at $T = 0K$, $\xi_0$ |
| The electron mean free path at $T = 4.2K$, $l_e$ |
| The residual resistance, $R_0$ |

The HEAT code solves the heat flow equations numerically from the interior wall to the exterior wall of a cavity at an accelerating gradient; and outputs the temperature distribution through the wall which is shown in Figure 6 [15].

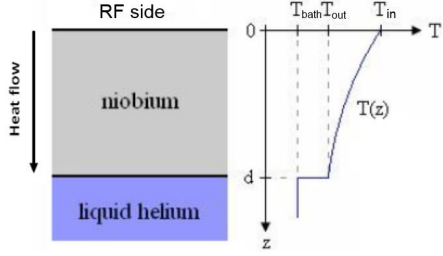

Figure 6: The sketch of the temperature distribution and the heat flow through a cavity wall.

The HEAT code adopts Koechlin and Bonin expression [18] to calculate niobium thermal conductivity. The Kapitza thermal conductivity is calculated from experimental data fitting [19]. The parameters which determined the thermal effects are listed in Table 2.

Table 2: The thermal parameters

| |
| --- |
| The wall thickness of the cavity, $d$ |
| The helium bath temperature, $T_{bath}$ |
| The phonon mean free path, $l_p$ |
| The residual-resistance ratio, $RRR$ |
| The accelerating gradient, $E_{acc}$ |
| The ratio of the peak magnetic field and the accelerating gradient, $H_p/E_{acc}$ |
| The geometry factor, $G$ |

In the HEAT-SRIMP fitting, the fitting parameters are listed in Table 3.

Table 3: The fitting parameters

| |
| --- |
| The transition temperature, $T_c$ |
| The energy gap at $T = 0K$ (entered as $\frac{\Delta}{kT_c}$); |
| The London penetration depth at $T = 0K$, $\lambda_0$ |
| The coherence length at $T = 0K$, $\xi_0$ |
| The electron mean free path at $T = 4.2K$, $l_e$ |
| The residual resistance, $R_0$ |

In an $R_s(T)$ fitting, the fitting parameters could be some of the parameters in Table 3, and the rest are set as fixed. The energy gap $\frac{\Delta}{kT_c}$, the electron-mean-free-path $l_e$, and the residual resistance $R_{res}$ are selected to be fit in most cases.

Therefore the form of the surface resistance is possible to be written as Eq. (5).

$$R_{s\_calc.} = f(T_{in}, P_{fix}, P_{fit}) \quad (5)$$

Here the constant $P_{fix}$ represents $f_0$ and the fixed parameters in Table 3; $P_{fit}$ is the fitting parameters listed in Table 3. From Eq. (3), the bath temperature can be expressed as a function of the inner temperature, the surface resistance, and the accelerating gradient in Eq. (6).

$$T_{bath} = g(T_{in}, R_s(T_{in}), E_{acc}, P_T) \quad (6)$$

Where the constant $P_T$ describes the parameters $d$, $l_p$, $RRR$, $H_p/E_{acc}$, and $G$ in Table 2. Here $T_{in}$ cannot be expressed as a single formula of $T_{bath}$, because $R_s$ is $T_{in}$ dependent as well. So we use an iterative method to solve Eq. (6) to obtain $T_{in}$, when cavity reaches thermal equilibrium. Then we can put $T_{in}$ into Eq. (5) to obtain $R_{s\_calc}$. It is clear that $R_{s\_calc}$ can be determined by all the parameters in Table 1, 2, and 3, which are given by Eq. (7).

$$R_{s\_calc.}(T_{bath}, E_{acc}, P_{fit}) = f(T_{bath}, E_{acc}, P_T, P_{fix}, P_{fit}) \quad (7)$$

The surface resistance from measurements is an array at a series of the bath temperatures $T_0$-$T_n$ as well as the accelerating gradients $E_0$-$E_m$, which is described in Eq. (8).

$$R_{s\_data}(T_{bath}, E_{acc}) = \begin{bmatrix} R_{00} & \cdots & R_{0m} \\ \vdots & \ddots & \vdots \\ R_{n0} & \cdots & R_{nm} \end{bmatrix} \begin{matrix} @T_0 \\ \vdots \\ @T_n \end{matrix} \quad (8)$$
$$@E_0 \cdots @E_m$$

The fitting program takes every $R_s(T_{bath})$ curves at different $E_{acc}$; compares Eq. (7) and Eq. (8); and tunes the parameter $P_{fit}$ to achieve the minimum fitting error by the least squares method. The fitting error $RSS$ is given by Eq. (9).

$$RSS = \sum_{i=1}^{n} \left( R_{s\_calc.}^i(T_{bath}^i, E_{acc}, P_{fit}) - R_{s\_data}^i(T_{bath}^i, E_{acc}) \right)^2 \quad (9)$$

The HEAT code and the SRIMP code are written in C++ and the least squares fitting program is written in Matlab. The Matlab program calls the C++ program as a function.

## COMPARISON BETWEEN HEAT-SRIMP FITTING AND SRIMP FITTING

*The parameters*

This section will give the comparison between the HEAT-SRIMP fitting and the SRIMP fitting. The purpose of the comparison is to demonstrate the fitting error by the simpler method. Here we use the HEAT-and-SRIMP program to generate the $Q_0(E_{acc})$ curves by the six cases with the parameters in Table 4. Then the $Q_0(E_{acc})$ curves are converted to the $R_s(T_{bath})$ curves. The correct method is ought to retrieve the parameters in Table 4 after fitting the curves.

Table 4: The parameters for generating the $Q_0(E_{acc})$ curves of the baked and the unbaked cases.

| Parameters | Baked | Unbaked |
|---|---|---|
| $l_e$ (A) | 300 | 6000 |
| $T_c$ (K) | 9.2 | |
| $\Delta/(kT_c)$ | 1.7-1.9 | |
| $\lambda_0$ (A) | 340 | |
| $\xi_0$ (A) | 640 | |
| $R_0$ (nΩ) | 5 | |
| $f_0$ (GHz) | 1.3 | |
| $d$ (mm) | 2.8 | |
| $l_p$ (μm) | 50 | |
| RRR | 300 | |
| $H_p/E_{acc}$ (mT/(MV/m)) | 4.2 | |
| $T_{bath}$ (K) | 1.45-2 | |
| $E_{acc}$ (MV/m) | 3-38 | |
| $G$ (Ω) | 278 | |

The fitting procedures of each case are same. Here we describe the fitting of the baked case with $\frac{\Delta}{kT_c} = 1.9$ in detail. Figure 7 is a series of $Q_0(E_{acc})$ curves under the bath temperature from 1.45K to 2K.

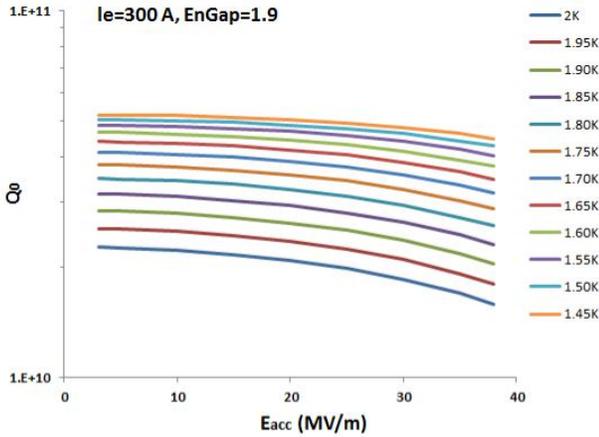

Figure 7: The $Q_0(E_{acc})$ curves of the baked case with $\frac{\Delta}{kT_c} = 1.9$ generated by the HEAT-and-SRIMP program.

Converting the $Q_0(E_{acc})$ curves to the $R_s(T_{bath})$ curves by Eq. (4), we obtain Figure 8. The curves have been fitted by the HEAT-SRIMP fitting program and the SRIMP fitting program respectively. In the fitting, the energy gap $\frac{\Delta}{kT_c}$ and the residual resistance $R_{res}$ have been selected as the fitting parameters. It has to be pointed out that the energy gap and the residual resistance is set-up as RF-field independent. The correct fitting method would be expected to retrieve the set-up value in Table 4. Table 5 and Table 6 give the input parameters of the HEAT-SRIMP fitting and the SRIMP fitting respectively.

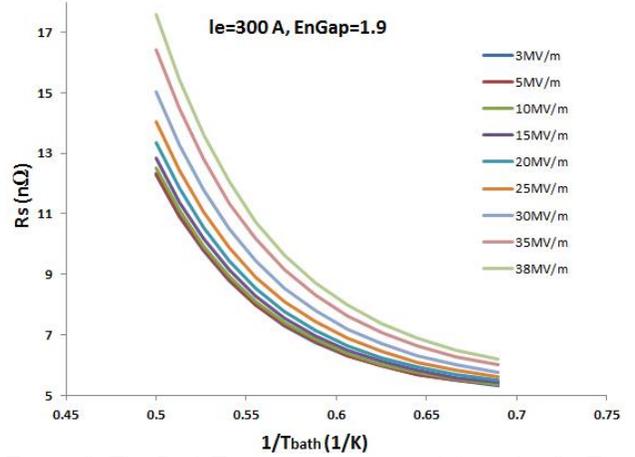

Figure 8: The $R_s$-$1/T_{bath}$ curves converted from the $Q_0$-$E_{acc}$ curves.

Table 5: The set-up parameters of the HEAT-SRIMP fitting program (The baked case)

| $T_c$ (K) | 9.2 |
|---|---|
| $\lambda_0$ (Å) | 340 |
| $\xi_0$ (Å) | 640 |
| $l_e$ (Å) | 300 |
| $f_0$ (GHz) | 1.3 |
| $d$ (mm) | 2.8 |
| $l_p$ (μm) | 50 |
| RRR | 300 |
| $H_p/E_{acc}$ (mT/(MV/m)) | 4.2 |
| $T_{bath}$ (K) | 1.45-2 |
| $E_{acc}$ (MV/m) | 3-38 |
| $G$ (Ω) | 278 |

Table 6: The set-up parameters of the SRIMP fitting program (The baked case)

| $T_c$ (K) | 9.2 |
|---|---|
| $\lambda_0$ (Å) | 340 |
| $\xi_0$ (Å) | 640 |
| $l_e$ (Å) | 300 |
| $f_0$ (GHz) | 1.3 |
| $T_{bath}$ (K) | 1.45-2 |
| $G$ (Ω) | 278 |

*The results*

The six cases have been calculated. It's very interesting to compare the two fitting methods in the baked case because these are the typical cases. The results are shown in Figure 9(a), (b), and (c). The results indicate the SRIMP fitting tends to give the energy gap value smaller than the correct value; and the SRIMP fitting has larger error in the lower energy gap case. The fitting error of the energy gap $\frac{\Delta}{kT_c}$ 1.9, 1.8, and 1.7 are 0.11, 0.17, and 0.27 respectively at 38 MV/m. The reason is the larger energy gap value gives the smaller BCS resistance; hence it has the smaller power losses, and thus the smaller thermal effects.

Figure 10(a), (b), and (c) exhibits the fitting results of the residual resistance. The SRIMP fitting results gave a larger residual resistance value than the expected value in all cases; the maximum error of the cases is around 0.4-0.8 nΩ shown in the figures. It is important that the trend of the SRIMP fitting error is not necessarily monotonic to the accelerating gradients. Every $R_s$-$1/T_{bath}$ curves at its $E_{acc}$ is independent. When the least squares program fits the energy gap and the residual resistance simultaneously, it searches the two parameters to achieve the minimum fitting errors.

It is very clear that the HEAT-SRIMP fitting extracts the correct values of the energy gap and the residual resistance for all the cases, which is showed in Figure 9 and Figure 10 as well.

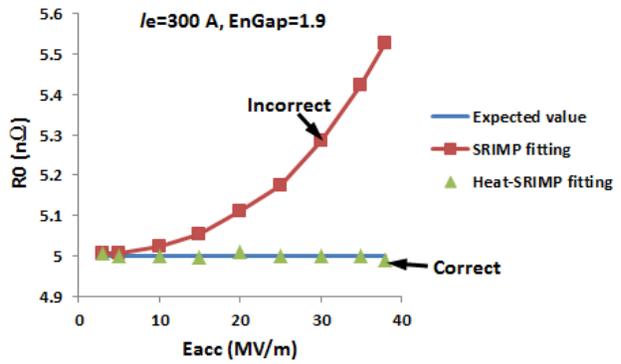
(a) The expected residual resistance value 5 nΩ with the energy-gap value 1.9.

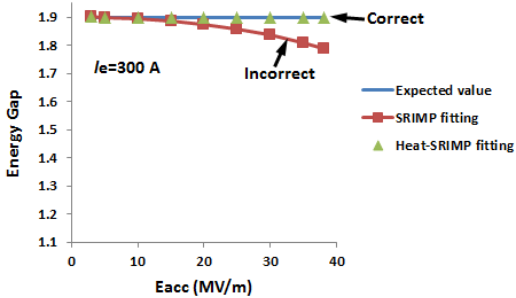
(a) The expected energy-gap value 1.9.

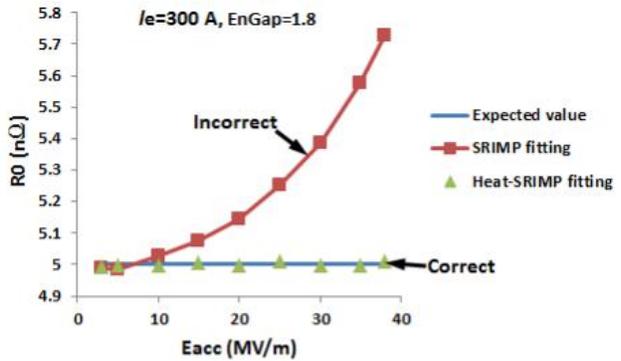
(b) The expected residual resistance value 5 nΩ with the energy-gap value 1.8.

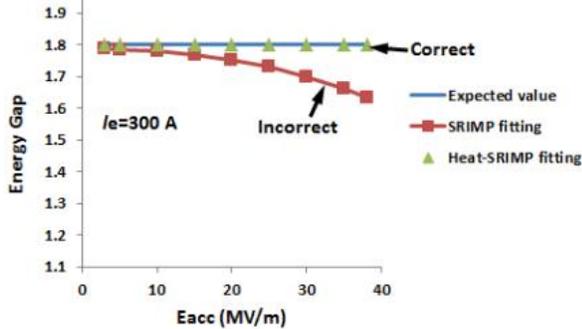
(b) The expected energy-gap value 1.8.

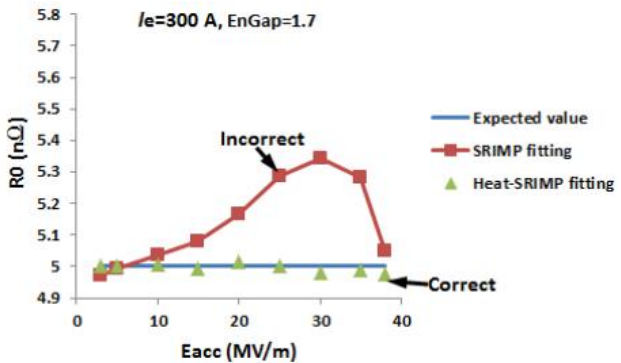
(c) The expected residual resistance value 5 nΩ with the energy-gap value 1.7.

Figure 10: The HEAT-SRIMP fitting and the SRIMP fitting comparison of the residual resistance of the baked case with the expected energy-gap value 1.7-1.9.

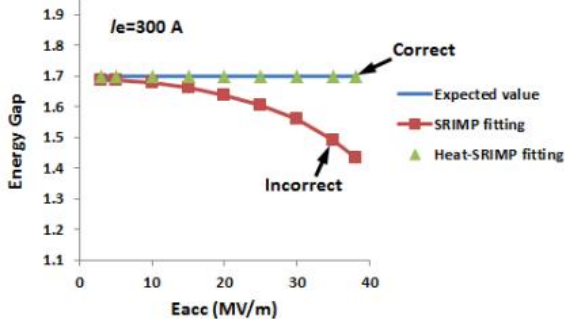
(c) The expected energy-gap value 1.7.

Figure 9: The HEAT-SRIMP fitting and the SRIMP fitting comparison of the baked case with the expected energy-gap value 1.7-1.9.

The unbaked cavities have the higher BCS resistance; so the Q-value is lower than the baked cavities. Therefore the thermal effects are reinforced in the unbaked cavities. Figure 11(a), (b), and (c) show the fitting results of the energy gap by the SRIMP fitting and the HEAT-SRIMP fitting with the expected value 1.9, 1.8, and 1.7 respectively. Even at low field ($E_{acc}$=3MV/m), the discrepancy of the SRIMP fitting and the correct value are obvious. Compared with the baked case, the fitting error is much larger in the unbaked case. But the trend of the fitting error versus the energy gap is same with the baked case.

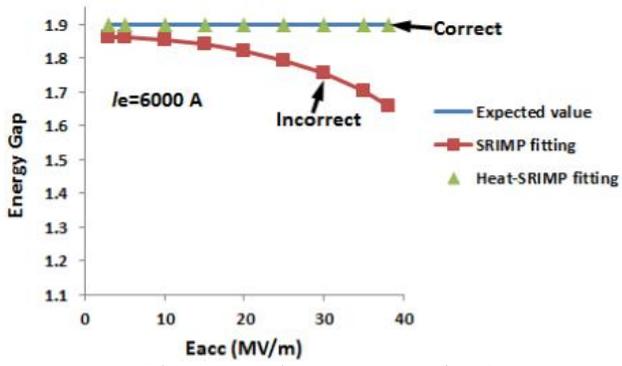
(a) The expected energy-gap value 1.9.

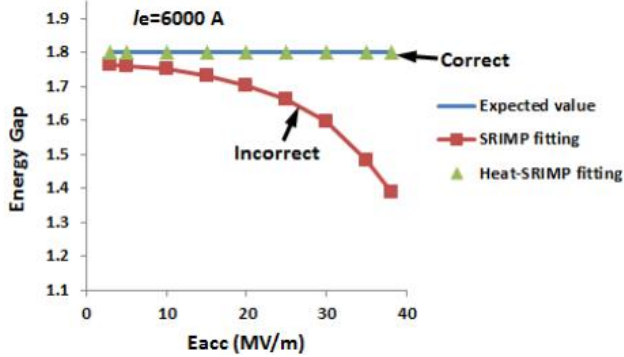
(b) The expected energy-gap value 1.8.

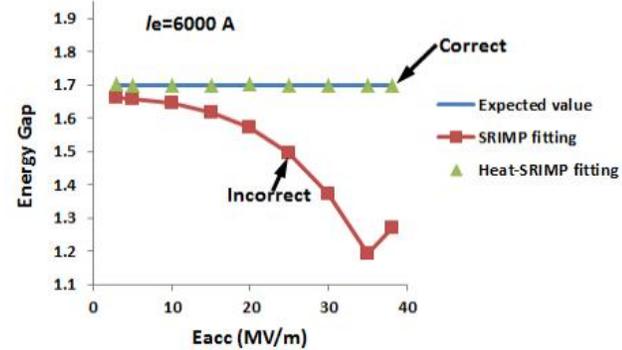
(c) The expected energy-gap value 1.7.

Figure 11: The HEAT-SRIMP fitting and the SRIMP fitting comparison of the un-baked case with the expected energy-gap value 1.7-1.9.

Figure 12 depicts the fitting results of the residual resistance by the SRIMP fitting and the HEAT-SRIMP fitting; (a), (b) and (c) show the result when the energy gap is 1.9, 1.8, and 1.7 respectively. The fitting error of the residual resistance in the SRIMP fitting is large; for example, in Figure 12(c) when the cavity gradient is at 38MV/m, the SRIMP fitting gave the residual resistance of 331 nΩ which is approximating 66 times larger than the correct value.

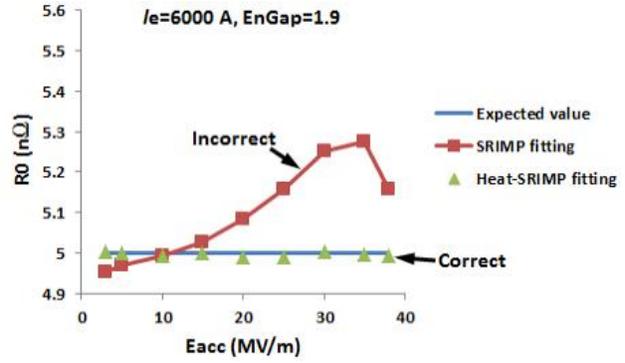
(a) The expected residual resistance value 5 nΩ with the energy-gap value 1.9.

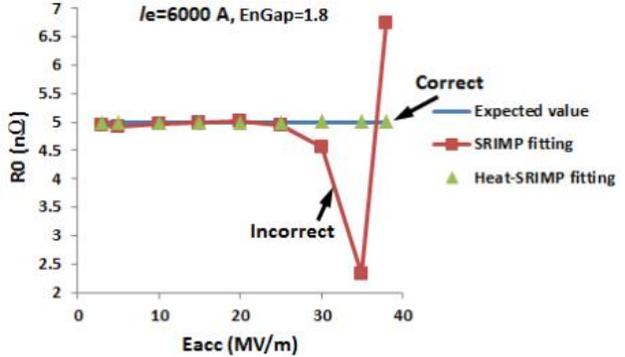
(b) The expected residual resistance value 5 nΩ with the energy-gap value 1.8.

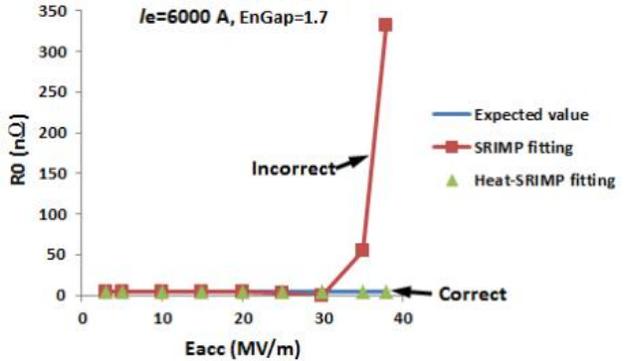
(c) The expected residual resistance value 5 nΩ with the energy-gap value 1.7.

Figure 12: The HEAT-SRIMP fitting and the SRIMP fitting comparison of the residual resistance of the un-baked case with the expected energy-gap value 1.7-1.9.

In the unbaked case, the HEAT-SRIMP fitting is able to extract the correct values of the energy gap and the residual resistance for all the cases, which is showed in Figure 11 and Figure 12 as well. It proves that the HEAT-SRIMP fitting successfully removed the thermal effects, hence eliminates the fitting error.

## CONCLUSION

This work demonstrated the valid method to extract the superconducting parameters from the $R_s(T_{bath})$ curves. The HEAR-SRIMP fitting program which is based on the thermal feedback model as well as the BCS theory has been developed. Compared to the SRIMP fitting program, it's able to remove the thermal effects caused by the thermal conductivity of Nb. Hence it eliminates the fitting error and provides the correct results.

# APPENDIX I: ERRORS OF USING THE GEOMETRY FACTOR FOR THE ELLIPTICAL CAVITIES

*Proof*

The magnetic-field $H$ distribution is un-uniform in elliptical-shape cavities *e.g.* TELSA cavities, so that the RF power dissipated on the inner surface is uneven because of the H-distribution and the inner-temperature distribution. However, are still can use the geometry factor $G$ to determine the surface resistance $R_s$ from the quality factor $Q_0$. In this section, we rigorously prove that the errors are tiny when $G$ is applied.

The quality factor $Q_0$ of a resonator which is the ratio of the stored energy and the power dissipated on the surface, per cycle, can be defined as Eq. (1):

$$Q_0 = \frac{\omega U}{P_c} = \omega \frac{\mu_0 \int_V H^2 dv}{\int_S R_s(T) H^2 ds} \tag{1}$$

Here $\omega = 2\pi f$ is the RF frequency of the resonator, $U$ is the stored energy, $P_c$ is the power loss on the surface. The integral in the numerator is taken over the volume of the cavity to calculate $U$; and the one in the denominator is taken over the surface of the cavity to calculate $P_c$. $H$ is the local magnetic field in the volume and on the surface respectively, and $T$ is the temperature on the local surface. $R_s(T)$ is the local surface resistance which is determined by $T$ on the surface. The geometry factor ($G$) is defined as Eq. (2):

$$G = \omega \frac{\mu_0 \int_V H^2 dv}{\int_S H^2 ds} \tag{2}$$

The surface resistance consists by the BCS resistance $R_{BCS}$ and the residual resistance $R_{res}$ showed in Eq. (3); and $R_{BCS}$ can be written by the approximate formula from the BCS theory in Eq. (4):

$$R_s(T) = R_{BCS}(T) + R_{res} \tag{3}$$

$$R_{BCS}(T) = A\left(\frac{1}{T}\right) f^2 \exp\left(-\frac{\Delta}{k_B T}\right) \tag{4}$$

From the thermal feedback model, it has been known that the temperature difference between $T_{in}$ and $T_{bath}$ is proportional to the power dissipated on the surface,

$$\Delta T = T_{in} - T_{bath} = \alpha P_{loss} = \alpha \frac{1}{2} R_s(T_{in}) H^2 \tag{5}$$

$$\alpha \equiv \frac{\kappa H_k}{\kappa + H_k d} \tag{6}$$

here the coefficient $\alpha$ is defined as Eq. (6), where $\kappa$ is the thermal conductivity of niobium, $H_k$ is the Kapitza conductance, $d$ is the thickness of cavity wall. Then we use Taylor series to expanse the BCS resistance in Eq. (4) at $T_{bath}$ to obtain the surface resistance of $T_{in}$ by Eq. (7):

$$R_s(T_{in}) = R_{BCS}(T_{bath}) + R'_{BCS}(T_{bath}) \Delta T + R_{res} \tag{7}$$

After applying Eq. (5) to Eq. (7), we obtain Eq. (8):

$$R_s(T_{in}) = R_s(T_{bath}) \left[1 - \frac{1}{2} R_{BCS}(T_{bath}) \left(\frac{\Delta}{k_B T_{bath}^2} - \frac{1}{T_{bath}}\right)\left(\frac{d}{\kappa} + \frac{1}{H_k}\right) H^2\right]^{-1} \tag{8}$$

then using $(1-x)^{-1} \approx 1 + x$, Eq. (8) can be written as Eq. (9):

$$R_s(T_{in}) = R_s(T_{bath}) \left[1 + \gamma \left(\frac{H}{H_c}\right)^2\right] \tag{9}$$

where $\gamma$ is

$$\gamma = \frac{1}{2} R_{BCS}(T_{bath}) H_c^2 \left(\frac{\Delta}{k_B T_{bath}^2} - \frac{1}{T_{bath}}\right)\left(\frac{d}{\kappa} + \frac{1}{H_k}\right) \tag{10}$$

Eq. (9) suggests that the surface resistance distribution caused by the inner temperature can be expressed by the distribution of the magnetic field. Hence the quality factor can be written as

$$Q_0 = \omega \frac{\mu_0 \int_V H^2 dv}{\int_S R_s(T_{in}) H^2 ds} = \omega \frac{\mu_0 \int_V H^2 dv}{\int_S R_s(T_{bath})\left(1 + \gamma \left(\frac{H}{H_c}\right)^2\right) H^2 ds} \tag{11}$$

Obviously the surface resistance is the function of *H*, and cannot be put out of the integral. But the magnetic field in the TESLA cavity is nearly flat in equator region and decays rapidly to iris region exhibited in Figure 1. Hence it can be approximately written as Eq. (12) which transforms the peak magnetic field ($H_p$) on surface into the surface resistance. Therefore the surface resistance is allowed to be extracted out of the integral, and the geometry factor *G* can be applied.

$$Q_0 = \omega \frac{\mu_0 \int_V H^2 dv}{R_s(T_{in})|_{H=H_p} \int_S H^2 ds} = \omega \frac{\mu_0 \int_V H^2 dv}{R_s(T_{bath})\left(1 + \gamma \left(\frac{H_p}{H_c}\right)^2\right)\int_S H^2 ds} = \frac{G}{R_s(T_{bath}, H_p)} \quad (12)$$

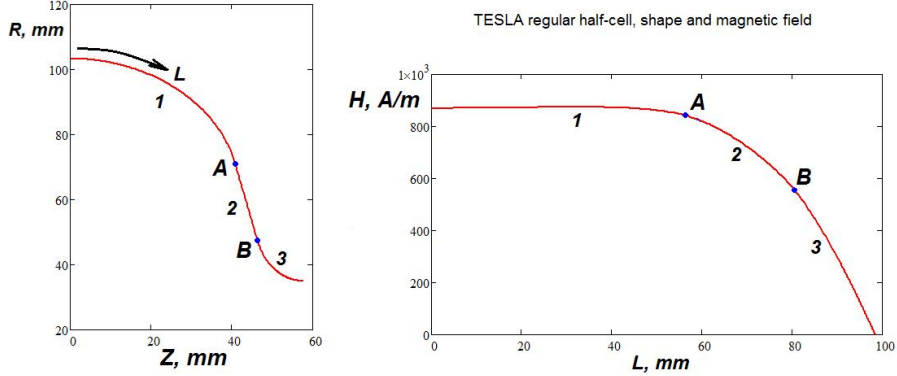

Figure 1: The TESLA 9-cell shape (left); and the surface magnetic-fields (right) calculated by the SLANS code.

It is important to understand the value of errors if we use Eq. (12) for the TESLA cavity. Here we use numerical way to calculate Eq. (11) and Eq (12). It's sufficient that we just compare the denominator parts of Eq. (11) and Eq. (12) which is written in Eq. (13) and Eq. (14) respectively.

$$\int_S R_s(T_{bath})(1 + \gamma H^2)H^2 ds$$
$$= 2\pi \sum_{i=0}^{218} R_s(T_{bath})\left[1 + \frac{\gamma}{H_c^2}\left(\frac{H_i + H_{i+1}}{2}\right)^2\right]\left[\left(\frac{H_i + H_{i+1}}{2}\right)^2 \left(\frac{R_i + R_{i+1}}{2}\right) dl\right] \quad (13)$$
$$= 16.317 \, (W)$$

$$R_s(T_{bath})(1 + \gamma H_p^2) \int_S H^2 ds$$
$$= 2\pi R_s(T_{bath})\left[1 + \gamma \left(\frac{H_p}{H_c}\right)^2\right]\sum_{i=0}^{218}\left[\left(\frac{H_i + H_{i+1}}{2}\right)^2 \left(\frac{R_i + R_{i+1}}{2}\right) dl\right] = 16.367 \, (W) \quad (14)$$

where

$$dl = \sqrt{(Z_i - Z_{i+1})^2 + (R_i - R_{i+1})^2} \quad (15)$$

The magnetic-field distribution is consisted by 218 discrete points, which are calculated at the accelerating gradient 38MV/m by the SLANS code [20]. $T_{bath}$=2K, $\gamma = 1$, $f_0$=1.3GHz, $R_s$(2K)=11.87nΩ, and $H_c$=2000Oe. It turns out that the error of Eq. (12) and Eq. (14) is 0.31% which is tiny. Table 1 is the summary of the errors of 1.3GHz elliptical cavities. The shapes and the field distribution of 1.3GHz elliptical cavities are showed in Figures 2-5.

Table 1: The errors of elliptical cavities

| Cavity shape | Errors (%) |
|---|---|
| TESLA 9-cell | 0.31 |
| TESLA single-cell | 0.30 |
| Re-entrant 9-cell | 0.15 |
| Cornell ERL 7-cell | 0.08 |
| Low Loss 9-cell | 0.17 |

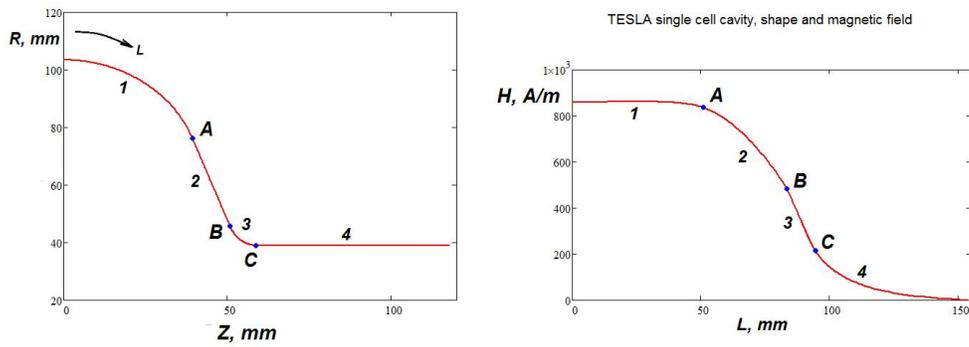
Figure 2: The TESLA single-cell shape (left); and the surface magnetic-fields (right) calculated by the SLANS code.

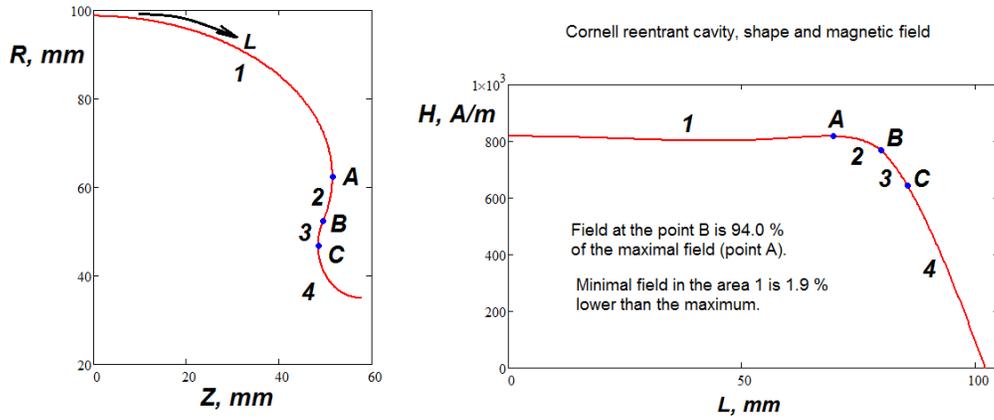
Figure 3: The Re-entrant 9-cell shape (left); and the surface magnetic-fields (right) calculated by the SLANS code.

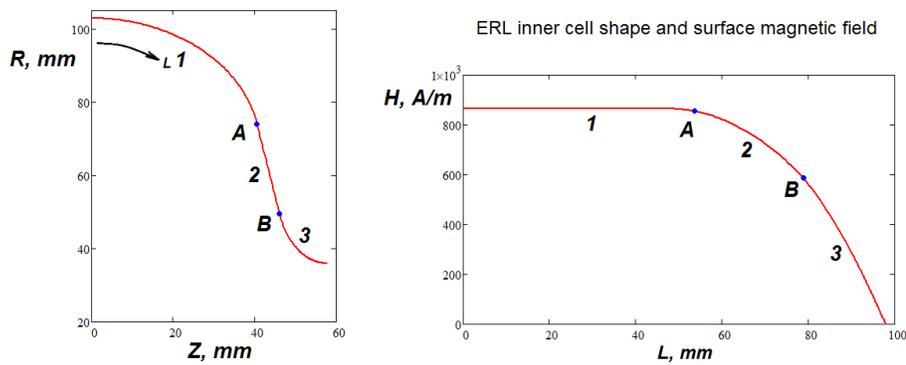
Figure 4: The Cornell ERL 7-cell shape (left); and the surface magnetic-fields (right) calculated by the SLANS code.

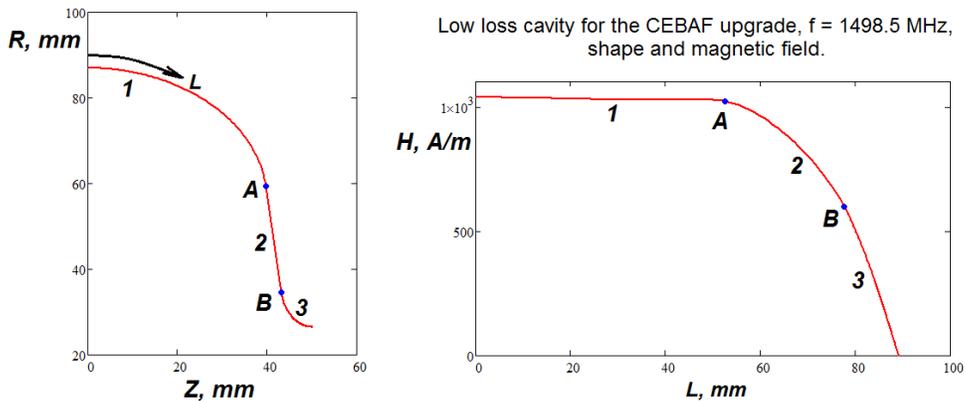
Figure 5: The Low Loss 9-cell shape (left); and the surface magnetic-fields (right) calculated by the SLANS code.